\documentclass[aps,prb,twocolumn,showpacs,groupedaddress]{revtex4-1}
\usepackage{amssymb}

\usepackage{graphicx}

\begin{document}

\title[]{Magnetoelectricity and Magnetostriction due to the Rare Earth Moment in TmAl$_3$(BO$_3$)$_4$}

\author{R. P. Chaudhury$^1$, B. Lorenz$^1$, Y. Y. Sun$^1$, L. N. Bezmaternykh$^2$, V. L. Temerov$^2$, and C. W. Chu$^{1,3}$}

\affiliation{$^1$ TCSUH and Department of Physics, University of Houston, Houston, Texas 77204-5002, USA}

\affiliation{$^2$ Institute of Physics, Siberian Division, Russian Academy of Sciences, Krasnoyarsk, 660036 Russia}

\affiliation{$^3$ Lawrence Berkeley National Laboratory, 1 Cyclotron Road, Berkeley, California 94720, USA}

\begin{abstract}
The magnetic properties, the magnetostriction, and the magnetoelectric effect in the d-electron free rare-earth aluminum borate TmAl$_3$(BO$_3$)$_4$ are investigated between room temperature and 2 K. The magnetic susceptibility reveals a strong anisotropy with the hexagonal c-axis as the hard magnetic axis. Magnetostriction measurements show a large effect of an in-plane field reducing both, the a- and c-axis lattice parameters. The magnetoelectric polarization change in a- and c-directions reaches up to 300 $\mu$C/m$^2$ at 70 kOe with the field applied along the a-axis. The magnetoelectric polarization is proportional to the lattice contraction in magnetic field. The results of this investigation prove the existence of a significant coupling between the rare earth magnetic moment and the lattice in $R$Al$_3$(BO$_3$)$_4$ compounds ($R$ = rare earth). They further show that the rare earth moment itself will generate a large magnetoelectric effect which makes it easier to study and to understand the origin of the magnetoelectric interaction in this class of materials.
\end{abstract}


\maketitle

Rare earth iron borates, RFe$_3$(BO$_3$)$_4$ (R=rare earth, Y), belonging to the trigonal system with space group R32\cite{joubert:68} have recently been studied because of
a large magnetoelectric effect observed in GdFe$_3$(BO$_3$)$_4$\cite{zvezdin:05,krotov:06,zvezdin:06,yen:06}, NdFe$_3$(BO$_3$)$_4$\cite{zvezdin:06b}, and
HoFe$_3$(BO$_3$)$_4$.\cite{chaudhury:09} Due to their noncentrosymmetric structure these materials possess also interesting optical properties which makes them potential
candidates for applications based on their good luminescent and nonlinear optical response.\cite{kalashnikova:04,gavriliuk:04,goldner:07} The iron-based compounds show a
wealth of magnetic phase transitions with the ordering of Fe-spins as well as the rare earth moments. The exchange interactions of the d-electron spins gives rise to
antiferromagnetic (AFM) order and the coupling of the d-electrons to the rare-earth moments with a strong single-ion anisotropy results in complex magnetic phases, spin reorientation phase
transitions, and collinear as well as non-collinear magnetic structures as resolved, for example, in neutron scattering experiments.\cite{fischer:06,ritter:07,ritter:08,mo:08}
The magnetoelectric interactions and the spontaneous electric polarization observed in some compounds \cite{chaudhury:09,popov:09} is intimately tied to the specifics of the magnetic orders. The complexity of the magnetic structures associated with the d- and f-electron moments makes it difficult to understand the microscopic details of the magnetoelectric effects observed in these materials.

Replacing the transition metal Fe with Al defines a new class of isostructural materials without d-electrons. $R$Al$_3$(BO$_3$)$_4$ ($R$ = rare earth) crystallize in the noncentrosymmetric structure, space group R32 (No. 155). The growth of single crystals of TmAl$_3$(BO$_3$)$_4$ and their structural properties have been described recently.\cite{jia:05} This material is of interest because of its nonlinear optical properties and the potential application as an infrared laser.\cite{jia:06} Accordingly, recent studies have focussed on optical absorption and emission properties of TmAl$_3$(BO$_3$)$_4$ single crystals.\cite{jia:06,malakhovskii:07}

Less attention was paid to the study of the dielectric and, possibly, magnetoelectric properties of TmAl$_3$(BO$_3$)$_4$. Due to the noncentrosymmetric structure a magnetoelectric coupling is not forbidden by symmetry, however, it is not clear whether or not the rare earth moment alone can give rise to a significant magnetoelectric effect. Unlike the $R$Fe$_3$(BO$_3$)$_4$ family of compounds, where a large magnetoelectric coupling has been observed for example in NdFe$_3$(BO$_3$)$_4$, possibly due to the d-electron spin of the Fe and its strong coupling to the lattice, the effects related to d-electron spins are completely missing in the rare earth aluminum borates. We have therefore investigated the magnetic and dielectric properties as well as the manetostriction and magnetoelectricity of TmAl$_3$(BO$_3$)$_4$ single crystals. The results show that a large magnetoelectric effect exists in compounds without d-electrons and that the rare earth magnetic moment may play a dominant role in the Al- as well as the Fe-based borates or even in the rare earth manganites.

Single crystals of TmAl$_3$(BO$_3$)$_4$ have been grown as described in Ref. \cite{temerov:08}. The crystals of about 10 mm size are transparent with a slightly green color and they show well defined smooth faces. Single crystal Laue X-ray spectroscopy was employed for orienting the crystals. They were cut and polished in different orientations according to the demands of the various experimental investigations. Measurements of the magnetization of TmAl$_3$(BO$_3$)$_4$ have been conducted along the hexagonal c- and a-axes in a 5 Tesla magnetometer (Magnetic Property Measurement System, Quantum Design). The magnetostriction was measured using the strain gauge method. The Wheatstone bridge was built from two gauges attached in parallel to the sample and two gauges glued to a piece of high purity copper serving as a reference sample. The imbalance of the bridge and the total bridge resistance was measured using the LR700 high resolution resistance bridge (Linear Research). The thermal expansion and magnetostriction was estimated upon changing temperature or magnetic field following standard procedures. The dielectric constant was derived from the capacitance of a sample shaped as a parallel plate capacitor with metallic contacts attached to two opposed surfaces. The high-precision capacitance bridge AH2500A (Andeen-Hagerling) was employed for the capacitance measurements. The magnetic field induced change of the electrical polarization was derived from integrating the magnetoelectric current measured by the K6517A electrometer (Keithley) while sweeping the magnetic field.

The magnetic susceptibility of TmAl$_3$(BO$_3$)$_4$ is highly anisotropic with the c-axis being the hard magnetic axis (Fig. 1). The magnetic anisotropy ($\chi_a$/$\chi_c$) increases with decreasing temperature and reaches a factor of $>$ 10 below 20 K. There is no indication of long-range magnetic order or a magnetic phase transition in the temperature range of 2 K $<$ T $<$ 300 K. The magnetic field dependence of the c-axis magnetization is perfectly linear up to the maximum field of 50 kOe as expected for a paramagnetic response. However, the a-axis magnetization shown in the inset to Fig. 1 is a nonlinear function of the field below about 40 K. The magnetically soft a-axis holds promise for a sizable magnetostrictive or magnetoelectric response when the field is applied in the hexagonal basal plane.

\begin{figure}
\begin{center}
\includegraphics[angle=0, width=2.5 in]{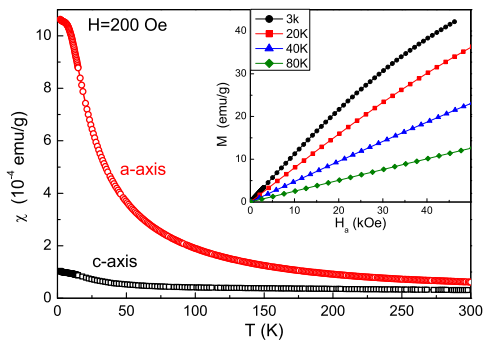}
\end{center}
\caption{(Color online) Magnetic susceptibility of TmAl$_3$(BO$_3$)$_4$ with the field oriented along the a- and c-axes. The inset shows the isothermal magnetization versus field measured along the a-axis.}
\end{figure}

The relative length change of both crystallographic orientations is displayed in Fig. 2. Near room temperature the thermal expansivities along a and c are 2.6x10$^{-6}$ K$^{-1}$ and 9.2x10$^{-6}$ K$^{-1}$, respectively. The c-axis length has a minimum near 50 K resulting in a negative expansivity at lower temperatures.

\begin{figure}
\begin{center}
\includegraphics[angle=0, width=2.5 in]{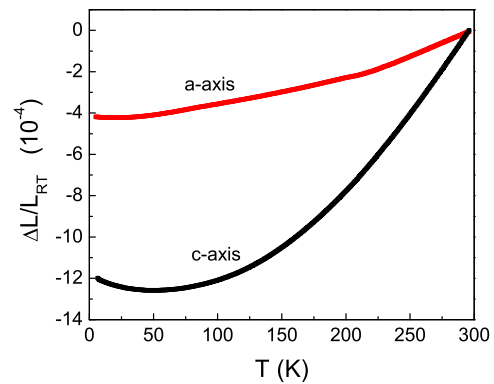}
\end{center}
\caption{(Color online) Change of lattice constants vs. temperature of TmAl$_3$(BO$_3$)$_4$.}
\end{figure}

Magnetostriction measurements with fields applied along the magnetically soft a-axis reveals a sizable change of both lattice constants, a and c. The relative length change of a and c is shown in Figs. 3a and 3b, respectively. At low magnetic field the lattice response is quadratic in the field as expected. The magnitude up to 80x10$^{-6}$ at 70 kOe and low temperature is large as compared to typical values for other magnetoelectric or multiferroic compounds. Typical values reported\cite{zvezdin:06} for BiFeO$_3$ (2x10$^{-6}$), GdMnO$_3$ (8x10$^{-6}$), GdFe$_3$(BO$_3$)$_4$ (10x10$^{-6}$), and NdFe$_3$(BO$_3$)$_4$ (30x10$^{-6}$) are significantly lower than the magnetostrictive coefficients of TmAl$_3$(BO$_3$)$_4$ (Fig. 3). It is also remarkable that a sizable longitudinal magnetostriction is still observed at relatively high temperatures, even above 100 K. The strong response of the lattice to the external magnetic field indicates an unusually large coupling of the Tm magnetic moments to the lattice.

Similar magnetostriction measurements carried out in magnetic fields along the c-axis show a much smaller effect. The maximum magnetostrictive strain in a c-axis field of 70 kOe is about 8x10$^{-6}$ at the lowest temperature of 3 K. The lower magnetostriction is consistent with the magnetic data showing that the c-axis is the hard magnetic axis.

\begin{figure}
\begin{center}
\includegraphics[angle=0, width=2.5 in]{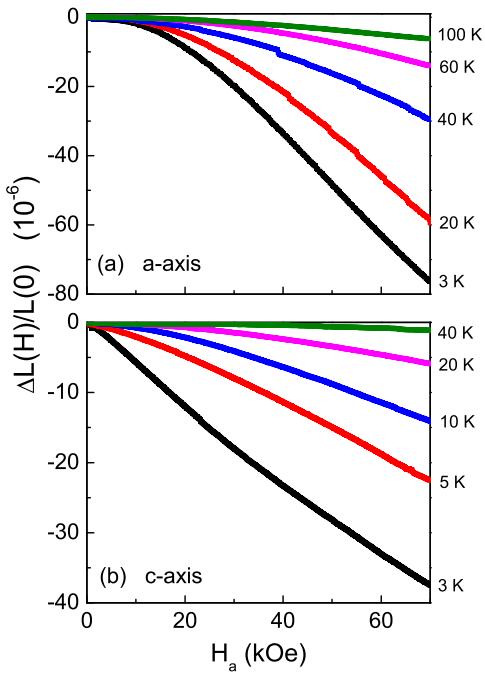}
\end{center}
\caption{(Color online) Longitudinal (a) and transverse (b) magnetostriction of TmAl$_3$(BO$_3$)$_4$ in magnetic fields applied along the a-axis.}
\end{figure}

The magnetic field induced change of the electric polarization is largest for magnetic fields oriented along the a-axis. Figure 4 shows the isothermal polarization change, $\Delta$P(H) at different temperatures. The data reveal a strong magnetoelectric effect with the polarization reaching 300 $\mu$C/m$^2$ at 3 K and 70 kOe. This value exceeds the magnitude of the field-induced polarization in GdFe$_3$(BO$_3$)$_4$ by a factor of 30 and it is comparable to the largest values observed in the Fe-based borate system, in NdFe$_3$(BO$_3$)$_4$, at the same magnetic field of 70 kOe.\cite{zvezdin:06} The result also indicates that the Fe d-electron spin is not necessary to achieve a large magnetoelectric effect in this class of compounds. The fact that the magnetoelectric polarization in many other $R$Fe$_3$(BO$_3$)$_4$ (e.g. $R$=Gd, Ho) is significantly smaller raises the question whether the Fe-spins facilitate or, possibly, oppose the magnetoelectric coupling in this class of compounds. The rare earth magnetic moment appears to play a primary role in the magnetoelectric coupling. It is also interesting to note that the 300 $\mu$C/m$^2$ exceeds the polarization values of many multiferroic compounds where the magnetic order actually induces a ferroelectric state, as for example in Ni$_3$V$_2$O$_8$ (P$_{max}$=170 $\mu$C/m$^2$)\cite{lawes:05}, MnWO$_4$ (P$_{max}$=70 $\mu$C/m$^2$)\cite{arkenbout:06}, or CuCrO$_2$ (P$_{max}$=120 $\mu$C/m$^2$)\cite{kimura:08b}.

\begin{figure}
\begin{center}
\includegraphics[angle=0, width=2.5 in]{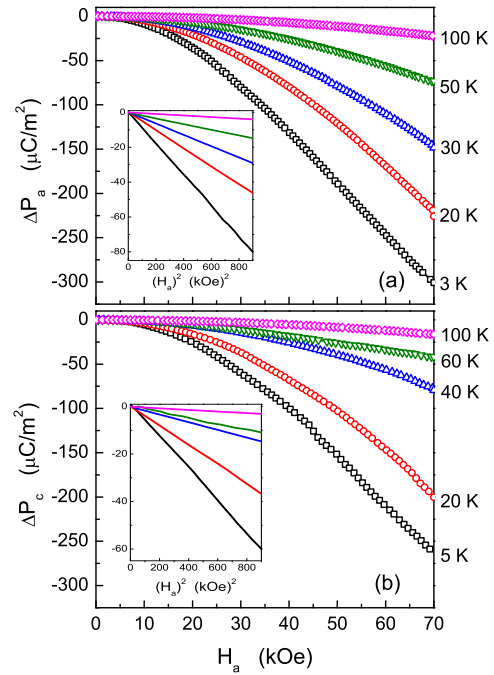}
\end{center}
\caption{(Color online) Longitudinal (a) and transverse (b) polarization change of TmAl$_3$(BO$_3$)$_4$ in magnetic fields applied along the a-axis. The insets show the quadratic field dependence of the polarization.}
\end{figure}

The increase of $\Delta$P(H) is nonlinear and it scales perfectly with H$^2$. Data up to 30 kOe are plotted as function of H$_a^2$ in the insets to Fig. 4 proving the second order character of the magnetoelectric coupling, $\Delta$P(H)$\sim$H$^2$. This strictly quadratic field dependence indicates that the linear magnetoelectric effect that may originate from the bi-linear term $\alpha_{ij}$E$_i$H$_j$ ($\alpha_{ij}$ linear magnetoelectric tensor) in the expansion of the free energy with respect to electric and magnetic fields\cite{fiebig:05} is apparently forbidden or negligibly small in TmAl$_3$(BO$_3$)$_4$. The higher order terms can dominate the magnetoelectric coupling\cite{eerenstein:06} as reported for paramagnetic piezoelectric crystals such as NiSO$_4\cdot$6H$_2$O.\cite{hou:65}

Magnetic fields applied along the c-axis were found far less effective in inducing a magnetoelectric polarization. The maximum value of $\Delta$P$_c$(H$_c$) of 10 $\mu$C/m$^2$ obtained at 70 kOe is more than an order of magnitude lower than the corresponding values in a-axis magnetic fields. This result is conceivable taking into account that the c-axis is the hard magnetic axis, the magnetization is less affected by the external field, and the longitudinal magnetostrictive strain is much smaller along the c-axis.

The magnetocapacitance (the change of the dielectric constant in magnetic fields) is found to be small. The maximum change of $\delta\varepsilon/\varepsilon$=$(\varepsilon(H)-\varepsilon(0))/\varepsilon(0)$ at low temperatures is 0.3 \% in a-axis magnetic fields up to 70 kOe.

\begin{figure}
\begin{center}
\includegraphics[angle=0, width=2.5 in]{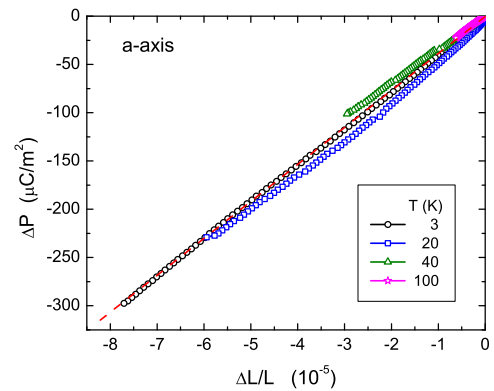}
\end{center}
\caption{(Color online) Longitudinal magnetoelectric polarization of TmAl$_3$(BO$_3$)$_4$ versus magnetostriction along the a-axis at different temperatures.}
\end{figure}

Comparing the magnetostriction data (Fig. 3) with the magnetoelectric polarization (Fig. 4) the similarity of their field dependence becomes obvious. A qualitative similarity of magnetostriction and magnetoelectric polarization was reported recently for GdFe$_3$(BO$_3$)$_4$ and NdFe$_3$(BO$_3$)$_4$.\cite{zvezdin:06} To verify the quantitative correlation between polarization and magnetostriction we plot the longitudinal a-axis polarization as a function of the a-axis strain induced by the field H$_a$. Fig. 5 shows data at different temperatures between 3 K and 100 K revealing a strictly linear correlation between both quantities. The slope, $\Delta$P/($\Delta$L/L)$\approx$4 C/m$^2$, is nearly independent of temperature which implies that the magnetoelectric polarization is mainly due to the field-induced lattice strain or magnetostriction. A significant piezoelectric effect is therefore expected in TmAl$_3$(BO$_3$)$_4$, but it is still awaiting experimental confirmation.

It is interesting to compare the current results for the d-electron free TmAl$_3$(BO$_3$)$_4$ with the extensive work published on $R$Fe$_3$(BO$_3$)$_4$. In the latter compounds the Fe-spins are strongly correlated and become magnetically ordered below temperatures of 40 K. A large magnetoelectric polarization was observed in NdFe$_3$(BO$_3$)$_4$.\cite{zvezdin:06b} The magnetic polarization of the rare earth moments in the internal field created by the antiferromagnetic order of the Fe-spins plays an important role in these compounds resulting in spin reorientation transitions, spontaneous polarization, and other novel phenomena, as for example in GdFe$_3$(BO$_3$)$_4$\cite{yen:06} and HoFe$_3$(BO$_3$)$_4$\cite{chaudhury:09}. The complexity of the magnetic and magnetoelectric interactions in these systems which consist of f- and d-electrons makes it very difficult to understand the physical origin of their magnetoelectric properties.

On the contrary, the absence of the d-electron spins in TmAl$_3$(BO$_3$)$_4$ makes it easier to investigate and understand the origin of the magnetoelectric effect. The Tm magnetic moments are only weakly correlated at high temperatures and they do not undergo magnetic long range order down to the lowest temperature (2 K) of this investigation. Nevertheless, a strong magnetoelectric polarization, comparable with the large values of NdFe$_3$(BO$_3$)$_4$, was detected. This result clearly shows that the Tm moment is strongly coupled to the lattice and it can give rise to a large magnetoelectric effect. Future investigations should focus on the magnetoelectric interactions in Al-based borates with different rare earth ions to further study the specifics of their magnetic moments, their interactions with the lattice, and the role of the single ion anisotropy.

In summary, we have shown that the d-electron free compound TmAl$_3$(BO$_3$)$_4$ gives rise to a large magnetoelectric effect that is comparable with the largest field-induced polarizations in the isostructural iron borate system, $R$Fe$_3$(BO$_3$)$_4$. The magnetoelectric polarization increases linearly with the magnetostrictive lattice contraction indicating the existence of a strong coupling of the rare earth magnetic moment with the lattice. The origin of the magnetoelectricity in TmAl$_3$(BO$_3$)$_4$ may be related to the piezoelectric effect of the noncentrosymmetric structure of the compound. Investigating the Al-based borate system sheds additional light onto the magnetoelectric properties of the related Fe-based class of compounds, $R$Fe$_3$(BO$_3$)$_4$, elucidating the dominant role of the rare earth magnetic moment in magnetoelectric borates.

\begin{acknowledgments}
This work is supported in part by the US Air Force Office of Scientific Research, the
T.L.L. Temple Foundation, the J. J. and R. Moores Endowment, and the
State of Texas through the TCSUH and at LBNL by the DoE.
\end{acknowledgments}

%Merlin.mbs v4.21 2009-07-09.

%\bibliography{HMO}

%

\end{document}